\pdfoutput=1
\documentclass[11pt]{article}

\usepackage[preprint]{acl}

\usepackage{listings}
\usepackage{times}
\usepackage{latexsym}
\usepackage[T1]{fontenc}
\usepackage[utf8]{inputenc}
\usepackage{microtype}
\usepackage{inconsolata}
\usepackage{graphicx}
\usepackage{booktabs}
\usepackage{multirow}
\usepackage{array}
\usepackage{colortbl}
\usepackage{makecell}
\usepackage[table]{xcolor}
\usepackage{amsmath,bm}
\usepackage{amssymb}
\usepackage{amsfonts}
\usepackage[hang,flushmargin]{footmisc} 
\usepackage[most]{tcolorbox}
\usepackage{listings}
\usepackage{multirow}
\usepackage{subcaption}
\usepackage{algorithm}
\usepackage{algorithmic}
\title{GLASS: GRPO-Trained LoRA for Acoustic Style Steering \\in Zero-Shot Text-to-Speech}

\author{Jaehoon Kang$^{1}$, Yejin Lee$^{1}$, Kyuhong Shim$^{1}$ \\
$^{1}$Department of Artificial Intelligence, Sungkyunkwan University, Korea \\
{\texttt{ \{morateng, yj.lee, khshim\}@skku.edu}}
}

\begin{document}
\maketitle

\begin{abstract}
We propose \textbf{GLASS}, a framework for composable acoustic style control in zero-shot autoregressive text-to-speech (TTS) that learns controls from post-generation rewards rather than style labels.
In zero-shot TTS, a speaker prompt often entangles speaker identity with prosodic attributes such as speaking rate and pitch, making it difficult to change style without changing the prompt itself.
GLASS instead treats each acoustic attribute as a reward-defined control direction.
For each control axis, GLASS freezes the TTS backbone and trains one lightweight LoRA adapter with Group Relative Policy Optimization (GRPO), using speech-token length and mean $F_0$ as style rewards and WER as an intelligibility anchor.
Because each control is represented as a LoRA weight update, independently trained adapters can be swapped, interpolated, and composed through linear LoRA arithmetic without retraining the backbone.
Experiments on speaking rate and pitch control show targeted style shifts while preserving naturalness, speaker similarity, and intelligibility, and demonstrate smooth interpolation and multi-axis composition across independently trained adapters.
\end{abstract}

\section{Introduction}\label{sec:intro}

Modern zero-shot text-to-speech (TTS) models~\citep{du2024cosyvoice,seedtts,f5tts,valle} can synthesize natural speech that closely matches the prompted speaker from a short reference prompt.
However, the same prompt also implicitly specifies prosody and acoustic style attributes such as speaking rate and pitch.
As a result, speaker identity and style are not independently addressable; changing the style of a speaker typically requires another prompt from the same speaker in the desired style. 
Furthermore, prompt selection offers only example-based control and provides no explicit mechanism for continuous interpolation or multi-axis composition.

A complementary line of controllable TTS methods conditions generation on style labels, acoustic attributes, or natural-language descriptions~\citep{stylespeech,prompttts,lyth2024natural}.
These controls are effective when the corresponding supervision is available, but adding new axes still requires annotations, reference examples, or curated data pipelines.
Moreover, such controls are often tied to predefined categories rather than continuous, speaker-preserving manipulation.

In this paper, we propose \textbf{GLASS}, a reward-guided training framework for composable acoustic style control in zero-shot TTS without requiring style annotations.
Given a fixed speaker prompt, GLASS aims to modify acoustic attributes while preserving speaker identity, intelligibility, and naturalness.
Instead of relying on target-style examples or style labels, GLASS defines each control axis through post-generation rewards.
Specifically, GLASS applies Group Relative Policy Optimization (GRPO)~\citep{deepseekmath,deepseekr1}: given multiple samples from the same text and speaker prompt, GLASS scores speech-token length for speaking rate, mean $F_0$ for pitch, and WER for intelligibility, then updates the model using relative within-group rewards.

To make these learned controls reusable and composable, GLASS stores each control axis as a lightweight LoRA adapter on a frozen backbone.
This keeps a shared backbone and one adapter per style direction; we show that independently trained controls can be swapped, interpolated, and composed at inference time via LoRA arithmetic~\citep{taskarithmetic,lorahub,ziplora}.
Training with multi-speaker prompts further encourages speaker-agnostic style directions that transfer to unseen prompts.

\begin{figure*}[t]
\centering
\includegraphics[width=\linewidth]{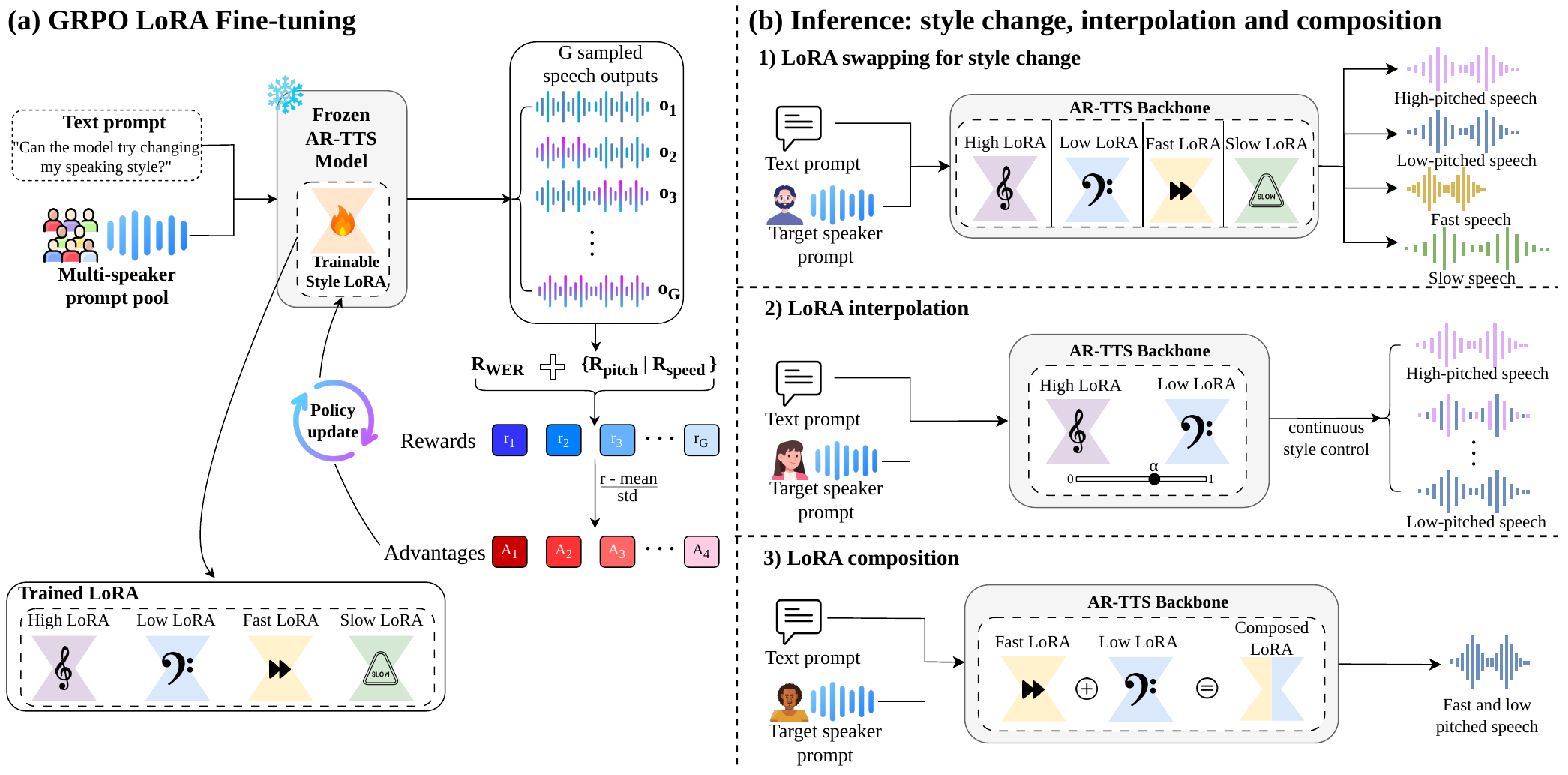}
\caption{GLASS trains LoRA adapters with GRPO using WER and acoustic style rewards, then combines the learned adapter pool for swapping, interpolation, and composition while keeping the speaker prompt fixed.}
\vspace{0.2cm}
\label{fig:overview}
\end{figure*}

Our contributions are as follows:
\vspace{-0.2cm}
\begin{itemize}
    \item We propose \textbf{GLASS}, a framework for composable acoustic style control in zero-shot TTS that modifies style while keeping the speaker prompt fixed, without requiring style labels.
    \vspace{-0.15cm}
    \item We introduce a GRPO-based objective that converts post-generation acoustic and ASR measurements into speaker-preserving TTS style directions.
    \vspace{-0.15cm}
    \item We show that the resulting independently trained LoRA adapters can be swapped, smoothly interpolated, and composed across axes via LoRA arithmetic, while preserving intelligibility, speaker similarity, and naturalness better than the baseline.

\end{itemize}
\section{Related Work}\label{sec:related}

\subsection{Style-controllable Text-to-Speech Models}

Style-controllable TTS has been studied through attribute conditioning, latent or reference style representations, natural-language prompts, and zero-shot speech prompts~\citep{ren2021fastspeech,lancucki2021fastpitch,wang2018style,zhang2019learning,prompttts,yang2024instructtts,valle,du2024cosyvoice,seedtts,f5tts,kang2026encoder,kang2026unlocking}.
These approaches provide effective control, but generally require style evidence in the form of predefined attributes, textual descriptions, or reference speech.
In contrast, GLASS learns style controls from measurable acoustic outcomes rather than from style annotations or reference examples.

\subsection{Reinforcement Learning in Text-to-Speech Models}

Preference optimization and reinforcement learning have recently been applied to TTS to optimize objectives beyond likelihood-based training.
For example, Seed-TTS uses WER and speaker-similarity rewards with REINFORCE~\citep{seedtts}, while subsequent GRPO-based TTS methods optimize ASR-, quality-, or prosody-related rewards~\citep{sun2025f5r,liu2026group,multirewardgrpo,dmospeech2}.
In parallel, Direct Preference Optimization (DPO)~\citep{rafailov2023direct} has been used with pairwise preferences to improve emotional control or prosodic naturalness when reliable automatic prosody rewards are unavailable~\citep{gao2025emo,shin2026no}.
These works primarily optimize the underlying generator or policy; GLASS is complementary, using GRPO to train lightweight LoRA adapters as modular style controls for swapping, interpolation, and composition.

\subsection{Adapter Arithmetic and Composable Control}

LoRA provides a parameter-efficient way to adapt large pretrained models by learning low-rank weight updates while keeping the base model frozen~\citep{hu2021lora}.
Beyond efficient fine-tuning, such weight updates can also be treated as modular control directions: task arithmetic composes model differences in weight space~\citep{taskarithmetic}, and LoRAHub combines multiple LoRA modules at inference time to adapt to unseen tasks without updating the backbone~\citep{lorahub}.
This adapter-arithmetic view has been especially visible in text-to-image diffusion, where independently trained LoRAs are merged or composed for subject, style, and multi-concept control~\citep{ziplora,gu2023mix,po2024orthogonal,gandikota2024concept}.

GLASS brings this perspective to acoustic style control in autoregressive zero-shot TTS.
Rather than training adapters from style labels, reference examples, or subject-specific data, GLASS learns speaker-agnostic LoRA directions from post-generation acoustic rewards such as speaking rate and mean $F_0$.
The resulting adapter library supports discrete swapping, continuous interpolation between opposite directions, and multi-axis composition while preserving the speaker prompt.

\section{Method}~\label{sec:method}
\vspace{-0.7cm}

\subsection{Backbone and Adapters}
We model autoregressive (AR) TTS as a speech-token policy $\pi(\mathbf{y}\mid\mathbf{x})$, where $\mathbf{x}$ contains text and a speaker prompt, and $\mathbf{y}$ is a sequence of discrete speech tokens.
GLASS freezes the base TTS model and learns one LoRA adapter per style direction $k$.

\begin{table*}[t]
\centering
\caption{Individual style control on Seed-TTS-eval \texttt{test\_en}. Syllables-per-second (SPS) and $F_0$ measure target styles; WER, SpkSim, UTMOS, S-MOS, and N-MOS measure quality. Bold highlights GLASS results in pairwise comparisons.}
\label{tab:main}
\footnotesize
\setlength{\tabcolsep}{3.0pt}
\renewcommand{\arraystretch}{1.05}
\begin{tabular*}{\textwidth}{@{\extracolsep{\fill}} l ccc ccccc @{}}
\toprule
& \multicolumn{3}{c}{\textbf{Style Metrics}} & \multicolumn{5}{c}{\textbf{Quality}} \\
\cmidrule(lr){2-4}\cmidrule(lr){5-9}
\textbf{Method} & SPS & $F_0$ M & $F_0$ F & WER\,$\downarrow$ & SpkSim\,$\uparrow$ & UTMOS\,$\uparrow$ & S-MOS\,$\uparrow$ & N-MOS\,$\uparrow$ \\
\midrule
Baseline (CosyVoice2-0.5B) & 3.65 & 120.4 & 192.2 & 2.81 & 0.655 & 3.28 & -- & -- \\
\rowcolor{gray!15}\multicolumn{9}{c}{\emph{Speed-axis control}} \\
\quad DSP speed-up   & 5.48 & 121.2 & 195.7 & 3.50 & 0.475 & 1.56 & 3.08 & 2.76 \\
\quad Fast LoRA (ours)             & \textbf{5.59} & 120.4 & 191.0 & 3.49 & \textbf{0.617} & \textbf{3.30} & \textbf{4.72} & \textbf{4.68} \\
\midrule
\quad DSP slow-down   & 2.19 & 121.7 & 193.7 & 2.67 & 0.500 & 1.45 & 2.76 & 2.28 \\
\quad Slow LoRA (ours)  & \textbf{2.30} & 122.1 & 194.3 & 3.18 & \textbf{0.650} & \textbf{3.05} & \textbf{4.56} & \textbf{4.24} \\
\rowcolor{gray!15}\multicolumn{9}{c}{\emph{Pitch-axis control}} \\
\quad DSP pitch-up     & 3.65 & 150.9 & 239.9 & 2.59 & 0.173 & 1.57 & 1.40 & 2.28 \\
\quad High-pitch LoRA (ours)       & 3.61 & \textbf{156.1} & \textbf{241.0} & 3.01 & \textbf{0.609} & \textbf{3.37} & \textbf{4.12} & \textbf{4.40} \\
\midrule
\quad DSP pitch-down     & 3.65 & 98.0  & 155.4 & 3.28 & 0.158 & 1.49 & 1.40 & 2.00 \\
\quad Low-pitch LoRA (ours)        & 3.68 & \textbf{108.9} & \textbf{164.6} & 3.11 & \textbf{0.632} & \textbf{3.16} & \textbf{4.60} & \textbf{4.84} \\
\bottomrule
\end{tabular*}
\end{table*}

\subsection{GRPO Training}
For each input $\mathbf{x}$, we sample a group of $G$ completions from the same text and speaker prompt, and assign a reward $r$ to each sample using an acoustic reward model $R_k$:
\begin{equation}
\mathbf{y}_1,\ldots,\mathbf{y}_G \sim \pi_{\theta_k}(\cdot\mid\mathbf{x}),\quad r_i = R_k(\mathbf{y}_i,\mathbf{x}).
\end{equation}
We convert these rewards into group-relative advantages by normalizing them within the sampled group:
\begin{equation}
A_i = \frac{r_i-\mu_r}{\sigma_r+\epsilon_{\mathrm{adv}}},
\qquad
\mu_r=\frac{1}{G}\sum_{j=1}^{G}r_j,
\end{equation}
where $\sigma_r$ is the within-group reward standard deviation.
This normalization makes the update depend on relative quality among samples from the same text and speaker prompt, rather than on the absolute reward scale.

Let $\ell_{\theta,i,t}=\log\pi_{\theta_k}(y_{i,t}\mid\mathbf{x},\mathbf{y}_{i,<t})$ be the current token log-probability for generated token $t$ in completion $i$.
We define $\ell_{\mathrm{old},i,t}$ analogously for the sampling policy and $\ell_{0,i,t}$ for the frozen reference policy, implemented by disabling LoRA layers in the same model.
The token-level likelihood ratio and its clipped version are
\begin{equation}
\begin{aligned}
\rho_{i,t}&=\exp(\ell_{\theta,i,t}-\ell_{\mathrm{old},i,t}),\\
\bar\rho_{i,t}&=\operatorname{clip}(\rho_{i,t},1-\varepsilon,1+\varepsilon).
\end{aligned}
\end{equation}
The sequence-level advantage $A_i$ is broadcast to all generated tokens, and losses are averaged within each generated sequence before averaging across completions.
With $\Delta_{i,t}=\ell_{0,i,t}-\ell_{\theta,i,t}$ and generated length $T_i$, the objective minimized for each LoRA adapter is
\begin{equation}
\begin{aligned}
\mathcal L_{\mathrm{GRPO}}
=&\frac{1}{G}\sum_{i=1}^{G}\frac{1}{T_i}\sum_{t=1}^{T_i}
\Big[-\min(\rho_{i,t}A_i,\bar\rho_{i,t}A_i)\\
&\quad+\beta\left(e^{\Delta_{i,t}}-\Delta_{i,t}-1\right)
\Big].
\end{aligned}
\end{equation}
The first term is the PPO-clipped policy-gradient loss, while the second term penalizes token-level drift from the frozen backbone.
Only LoRA parameters receive gradients; all non-LoRA TTS components remain frozen.

Because rewards are computed after waveform generation, the same loop can optimize non-differentiable signals such as ASR-based WER, generated token length, and estimated $F_0$.

\subsection{Reward Design}
Each reward combines a WER intelligibility anchor with a style term:
\begin{equation}
R_k(\mathbf{y},\mathbf{x}) = \eta R_{\mathrm{WER}}(\mathbf{y},\mathbf{x}) + (1-\eta)R^{\mathrm{style}}_k(\mathbf{y}),
\end{equation}

where $R_{\mathrm{WER}} = 1 - \tanh(\gamma\,\mathrm{WER}(\mathbf{x},\mathbf{y}))$ follows prior WER-based TTS rewards~\citep{shin2026no}; WER is computed from Whisper transcripts~\citep{radford2023robust}.
For style, we min--max normalize the target statistic $z_i$ within each GRPO group:
\begin{equation}
m(z_i)=
\begin{cases}
\dfrac{z_i-z_{\min}}{z_{\max}-z_{\min}}, & z_{\max}>z_{\min},\\[0.4em]
0.5, & \text{otherwise}.
\end{cases}
\end{equation}
Here $z_{\min}$ and $z_{\max}$ are group extrema, and $0.5$ is used when all completions tie.
Speed rewards use generated speech-token length, with fast and slow taking $1-m(z_i)$ and $m(z_i)$, respectively; pitch rewards use utterance-level mean $F_{0,i}$, with high and low taking $m(z_i)$ and $1-m(z_i)$, respectively.

\subsection{LoRA Arithmetic}
A LoRA adapter represents a low-rank update $\Delta W_k$ to frozen weights.
Each adapter learns one signed acoustic direction while the speaker prompt and backbone remain fixed.
Since all adapters share target modules and rank, their updates can be combined directly at inference time:
\begin{equation}
\Delta W(\mathbf{w}) = \sum_k w_k \Delta W_k.
\end{equation}
The weights $w_k$ are user-tunable controls, enabling same-axis interpolation such as $\Delta W(\alpha)=\alpha\Delta W_{\mathrm{fast}}+(1-\alpha)\Delta W_{\mathrm{slow}}$, and multi-axis composition such as $\Delta W=w_1\Delta W_{\mathrm{fast}}+w_2\Delta W_{\mathrm{high}}$ without retraining.
We use $w{=}0.5$ as a stable composition point to avoid full-strength overshoot.

\section{Experiments}
\label{sec:experiments}

\begin{figure*}[t]
\centering
\includegraphics[width=\linewidth]{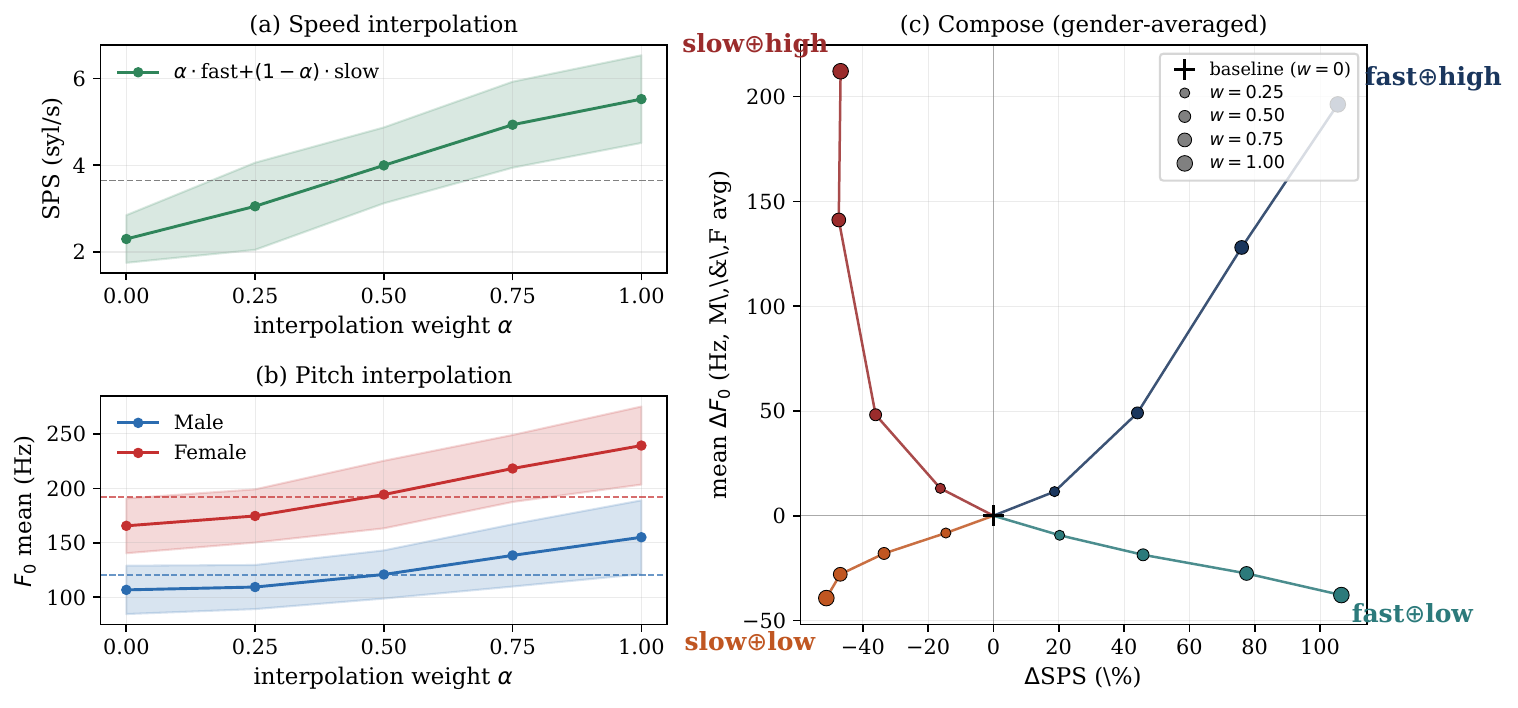}
\caption{Continuous interpolation and gender-averaged multi-axis composition via linear LoRA arithmetic. (a,b) Opposite-adapter interpolation smoothly changes SPS and $F_0$. (c) Speed--pitch composition moves outputs toward intended quadrants; $w{=}0.5$ is stable before larger pitch overshoot.}
\label{fig:interp_comp}
\vspace{0.3cm}
\end{figure*}

\begin{table*}[t]
\centering
\caption{Continuous interpolation between opposite LoRA adapters on the 200-utterance subset. The target style metric ($\alpha$ vs.\ SPS for speed; vs.\ $F_0$ for pitch) varies smoothly while off-axis metrics remain near baseline. WER reaches its minimum at the central blend $\alpha{=}0.5$ on both axes.}
\label{tab:interp_quality}
\footnotesize
\setlength{\tabcolsep}{2.0pt}
\begin{tabular*}{\textwidth}{@{\extracolsep{\fill}}lccccc|ccccc@{}}
\toprule
& \multicolumn{5}{c|}{\textit{Speed axis} ($\alpha\cdot$fast$+(1{-}\alpha)\cdot$slow)} & \multicolumn{5}{c}{\textit{Pitch axis} ($\alpha\cdot$high$+(1{-}\alpha)\cdot$low)} \\
\midrule
$\alpha$ & SPS & WER\% $\downarrow$ & SpkSim $\uparrow$ & UTMOS $\uparrow$ & $F_0$ (M/F) & SPS & WER\% $\downarrow$ & SpkSim $\uparrow$ & UTMOS $\uparrow$ & $F_0$ (M/F) \\
\midrule
0.00 & 2.30 & 4.50 & 0.647 & 3.00 & 123/196 & 3.72 & 3.36 & 0.629 & 3.16 & 107/166 \\
0.25 & 3.05 & 4.25 & 0.649 & 3.18 & 121/193 & 3.71 & 2.60 & 0.644 & 3.19 & 109/175 \\
0.50 & 4.00 & \textbf{2.16} & 0.645 & 3.29 & 121/194 & 3.64 & \textbf{2.14} & 0.644 & 3.28 & 121/194 \\
0.75 & 4.93 & 2.91 & 0.624 & 3.32 & 121/193 & 3.64 & 3.02 & 0.634 & 3.34 & 138/218 \\
1.00 & 5.52 & 3.51 & 0.614 & 3.31 & 119/192 & 3.62 & 2.96 & 0.614 & 3.35 & 155/239 \\
\bottomrule
\end{tabular*}
\end{table*}

\subsection{Setup}
\label{sec:setup}

We train one LoRA adapter per style direction on LibriTTS-R~\cite{koizumi2023libritts} using multi-speaker prompts.
We evaluate zero-shot transfer on Seed-TTS-eval \texttt{test\_en} ($N{=}1088$), reporting style-control metrics (syllables per second (SPS) and $F_0$) and quality metrics (WER, speaker similarity (SpkSim), UTMOSv2, S-MOS, and N-MOS).
This setup tests whether reward-trained LoRA adapters behave as speaker-agnostic style directions that transfer to unseen prompts, rather than memorizing training speakers.
Baselines are the unmodified CosyVoice2-0.5B~\cite{du2024cosyvoice} model and digital signal processing (DSP)-based transformations of its outputs: time-stretching for speed and pitch-shifting for pitch.

In our implementation, LoRA is attached only to the autoregressive token generation module, yielding 1.08M trainable parameters, or 0.22\% of the 495M-parameter AR backbone.
Figure~\ref{fig:overview} summarizes the training and inference workflow.
Additional details on implementation, metrics, and data subsets are provided in Appendix~\ref{app:details}.

\subsection{Individual style control}
\label{sec:individual}

Table~\ref{tab:main} shows that LoRA adapters reach DSP-comparable target shifts (SPS, $F_0$) while keeping off-axis statistics near the baseline, indicating largely separable speed and pitch directions.
In particular, DSP baselines collapse UTMOS to 1.45--1.57 and, for pitch shifting, drop SpkSim below 0.18, whereas LoRA stays within 0.23 UTMOS and 0.05 SpkSim of the baseline.
S-MOS/N-MOS also favor LoRA over DSP (4.12--4.72/4.24--4.84 vs. 1.40--3.08/2.00--2.76).
WER stays within 0.7 pp of the baseline across all adapters.

\subsection{Continuous interpolation}
\label{sec:interp}

We linearly blend opposite adapters with $\alpha\in\{0,0.25,0.5,0.75,1.0\}$ on the gender-balanced 200-utterance subset.
Figure~\ref{fig:interp_comp}(a,b) shows smooth control: SPS shifts from 2.30 to 5.52, while male/female $F_0$ moves from 107/166 Hz to 155/239 Hz.
Table~\ref{tab:interp_quality} reports the same sweep together with quality metrics.
WER follows a U-shape with a minimum at the central blend $\alpha{=}0.5$ on both axes (2.16\% for speed and 2.14\% for pitch), below both endpoints.
UTMOSv2 remains within 0.35 of the baseline, and SpkSim is nearly flat for $\alpha\le0.5$ before dropping modestly at the most extreme settings.
Together with the utterance-level monotonicity analysis in Appendix~\ref{app:interp}, these results show that interpolation provides a stable continuous control regime rather than a degenerate average of two adapters.

\subsection{Multi-axis composition}
\label{sec:comp}

We then compose one speed adapter and one pitch adapter across four combinations: fast$\oplus$high, fast$\oplus$low, slow$\oplus$high, and slow$\oplus$low.
At $w_A{=}w_B{=}0.5$, all four combinations move toward their intended speed--pitch quadrant (Figure~\ref{fig:interp_comp}(c)), with each axis retaining 80--121\% of its single-axis effect (Appendix~\ref{app:compose}, Table~\ref{tab:compose_retention}).
Appendix~\ref{app:compose} also provides gender-specific views and shows why full-strength composition overshoots.
The result demonstrates that independently reward-trained adapters can be reused as composable acoustic style directions.

\section{Conclusion and Future Work}\label{sec:conclusion}

We presented GLASS, a reward-guided framework for modular acoustic style control in AR-TTS.
By freezing the TTS backbone and training one lightweight LoRA adapter per style direction with GRPO, GLASS learns speaker-agnostic speed and pitch controls from post-generation rewards rather than style-labeled speech.
Experiments show that the resulting adapters modify speaking rate and pitch while preserving intelligibility and speaker similarity.
Since styles are encoded as LoRA weight updates, the same adapters support swapping, smooth interpolation, and multi-axis composition, suggesting a practical path toward extensible and composable style control for zero-shot TTS.
More broadly, GLASS points toward scalable libraries of reward-trained style adapters.
Future work will extend GLASS to richer style attributes and robust adapter-weight selection.

\newpage
\section*{Limitations}

This work focuses on two measurable acoustic style axes, speaking rate and pitch. These axes are interpretable and useful, but they do not cover richer prosodic or paralinguistic attributes such as emotion, speaking effort, accent, or conversational style. Extending GLASS to such attributes would require reliable automatic rewards or human preference feedback.

Our rewards and evaluations rely on automatic proxy measurements: speech-token length for speaking rate, voiced-frame mean $F_0$ for pitch, Whisper-based WER for intelligibility, and automatic speaker-similarity and naturalness metrics. These proxies make the method reproducible and scalable, but they may not capture all perceptual aspects of style control.

% \section*{Acknowledgments}
% Ack

\bibliography{custom}

\newpage
\appendix
\section{Implementation and Evaluation Details}
\label{app:details}

\paragraph{Backbone and LoRA.}
The backbone is CosyVoice2-0.5B.
LoRA is attached only to the Qwen2 autoregressive token model, while speech embeddings, token decoder, flow-matching acoustic model, and vocoder remain frozen.
We target $\{q_{\mathrm{proj}},v_{\mathrm{proj}}\}$ attention projections with rank $r{=}16$, scaling $\alpha{=}32$, and dropout $0.05$.

\paragraph{Training data and optimization.}
For each style adapter, we sample prompts from 50 randomly selected LibriTTS-R~\citep{koizumi2023libritts} \texttt{train-clean-100} speakers (25 male, 25 female; seed 42), each with at least 20 utterances.
Target texts are drawn from a 3{,}000-sentence pool from the same corpus.
For each batch item, we sample a fresh speaker uniformly and keep that speaker fixed across the $G{=}8$ generations in the GRPO group, making within-group reward normalization comparable under the same text and speaker condition.
This multi-speaker prompt training encourages each LoRA to capture a transferable style direction rather than a speaker-specific voice transformation.

Each adapter is trained for 500--750 update steps with AdamW, batch size 4, group size $G{=}8$, two PPO epochs, $\varepsilon{=}0.2$, $\beta{=}0.01$, $\eta{=}0.5$, and $\gamma{=}1$.
Whisper-large-v3~\citep{radford2023robust} provides transcripts for $R_{\mathrm{WER}}$; token length is computed from generated speech tokens, and pitch rewards use voiced-frame $F_0$ estimated with \texttt{pyworld}.

\paragraph{Evaluation.}
Seed-TTS-eval \texttt{test\_en} contains 1{,}088 prompt--text pairs from Common Voice and is out-of-domain with respect to the LibriTTS-R training data.
This setting tests whether LoRA style adapters trained with LibriTTS-R speakers transfer to unseen speaker prompts at inference.
For interpolation sweeps, we use a gender-balanced 200-utterance subset (100 male, 100 female) to keep inference cost tractable.

We report SPS, mean voiced-frame $F_0$, WER, speaker similarity, UTMOSv2~\citep{saeki2022utmos}, S-MOS, and N-MOS.
SpkSim is the cosine similarity between prompt and generated speech embeddings extracted by a WavLM-large speaker-verification model~\citep{chen2022wavlm}, following the Seed-TTS evaluation protocol.
For MOS, 15 human raters scored 25 utterances per transformed system on 1--5 S-MOS/N-MOS scales for prompt speaker similarity and naturalness; each utterance received five ratings, and Table~\ref{tab:main} reports means.
Following Spark-TTS~\citep{wang2025spark}, SPS denotes syllables per second, computed as the target-text syllable count divided by the generated waveform duration in seconds; higher SPS indicates faster speaking rate.
For DSP baselines, we apply \texttt{librosa} time-stretching with rates $1.5$ and $0.6$ for speed and pitch shifts of $\pm4$ semitones for pitch.

\section{Continuous Interpolation Details}
\label{app:interp}

We linearly blend opposite adapters as $\Delta W(\alpha) = \alpha\,\Delta W_{+} + (1-\alpha)\,\Delta W_{-}$ with $\alpha \in \{0,0.25,0.5,0.75,1.0\}$, where $+$ denotes the fast or high-pitch adapter and $-$ denotes its opposite. For tractability, we evaluate on a gender-balanced 200-utterance subset of Seed-TTS-eval \texttt{test\_en} (100\,M, 100\,F; seed 42). The same prompts and texts are used across all $\alpha$, so trends are within-utterance.

\paragraph{Monotonicity.} We compute per-utterance Spearman $\rho$ between $\alpha$ and the target acoustic statistic (SPS for speed; mean voiced-frame $F_0$ for pitch, restricted to utterances with voicing ratio $>0.3$), then average across utterances. The averaged correlations are strongly positive: $0.954{\pm}0.088$ for speed, $0.874{\pm}0.154$ for male pitch, and $0.923{\pm}0.099$ for female pitch. Thus, the target statistic usually increases with $\alpha$ at the utterance level; the curves in Figure~\ref{fig:interp_comp}(a,b) are not artifacts of averaging.

\section{Composition Retention and Overshoot}
\label{app:compose}

We compose one speed adapter and one pitch adapter via linear LoRA arithmetic $\Delta W = w_A\,\Delta W_A + w_B\,\Delta W_B$ with $w_A{=}w_B{=}w$ for $w\in\{0.25,0.5,0.75,1.0\}$. We evaluate these compositions on the full Seed-TTS-eval \texttt{test\_en} ($N{=}1088$).

\begin{table}[t]
\centering
\caption{Per-axis retention at $w_A{=}w_B{=}0.5$. Both axes recover 80--121\% of the single-axis effect, indicating near-additive composition.}
\label{tab:compose_retention}
\small
\begin{tabular*}{\columnwidth}{@{\extracolsep{\fill}}lccc@{}}
\toprule
Combination & ret$_{\text{SPS}}$ & ret$_{F_0,\text{M}}$ & ret$_{F_0,\text{F}}$ \\
\midrule
fast$\oplus$high & 83\% & 121\% & 114\% \\
fast$\oplus$low  & 86\% &  90\% &  91\% \\
slow$\oplus$high & 98\% & 109\% & 117\% \\
slow$\oplus$low  & 91\% &  80\% &  82\% \\
\bottomrule
\end{tabular*}
\end{table}

\begin{figure}[t]
\centering
\includegraphics[width=\columnwidth]{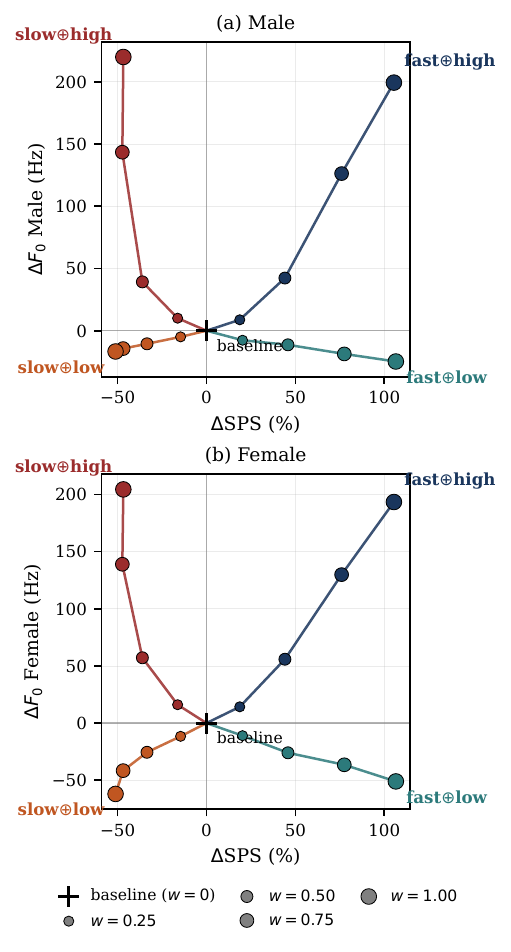}
\caption{Gender-specific multi-axis speed--pitch composition results corresponding to the averaged view in Figure~\ref{fig:interp_comp}(c). Both male and female subsets move toward the intended quadrants as $w$ increases.}
\label{fig:composition_gender}
\end{figure}

\paragraph{Retention at $w{=}0.5$.}
To quantify whether the composed adapter preserves each axis's effect, we define
\[
\text{ret}_X \;=\; \frac{\Delta X_{\,\text{compose}(w{=}0.5)}}{\Delta X_{\,\text{single-axis LoRA}}} \times 100\%,
\]
where $\Delta X$ is the signed change in target statistic $X$ (SPS \% or $F_0$ in Hz) relative to the baseline. Table~\ref{tab:compose_retention} shows that at $w{=}0.5$, both axes retain 80--121\% of their respective single-axis effect. Retention is consistently close to and occasionally above 100\%, suggesting that the two adapters act on largely separable directions in token-policy space rather than canceling each other. Mild overshoot ($>$100\%) appears on the pitch axis when paired with \texttt{high}, which may reflect longer high-pitched intervals contributing more voiced frames to the $F_0$ estimate.

\begin{table}[t]
\centering
\caption{Composition at $w_A{=}w_B{=}1.0$. Pitch overshoots to non-physiological registers when paired with \texttt{high} (male $F_0$ up to $5.6\times$ the single-axis shift); voicing ratio drops sharply when paired with \texttt{low}, indicating creak/aperiodic phonation.}
\label{tab:compose_overshoot}
\small
\begin{tabular*}{\columnwidth}{@{\extracolsep{\fill}}lcccc@{}}
\toprule
Combination & SPS & $F_0$(M) & $F_0$(F) & voicing \\
\midrule
fast$\oplus$high & 7.50 & 318.9 & 385.5 & 0.64 \\
fast$\oplus$low  & 7.54 & 106.2 & 147.0 & 0.37 \\
slow$\oplus$high & 1.94 & 326.3 & 392.0 & 0.58 \\
slow$\oplus$low  & 1.78 & 106.8 & 136.1 & 0.32 \\
\midrule
Baseline         & 3.65 & 120.4 & 192.2 & 0.65 \\
\bottomrule
\end{tabular*}
\end{table}

\paragraph{Why $w{=}0.5$, not $w{=}1.0$.}
While naive intuition suggests using each adapter at its trained strength ($w{=}1.0$), summing two adapters at full strength produces pronounced overshoot, especially on the pitch axis. Table~\ref{tab:compose_overshoot} reports the $w{=}1.0{+}1.0$ regime: fast$\oplus$high drives male mean $F_0$ from 120\,Hz at baseline to 319\,Hz---roughly $5.6\times$ the shift produced by the single-axis high adapter (35.7\,Hz), entering an unrealistically high register. Female mean $F_0$ similarly reaches 386\,Hz. The combinations involving \texttt{low} show a different failure mode: mean voicing ratio collapses to $0.32$--$0.37$ (baseline $\approx0.65$), indicating creaky or aperiodic phonation. We therefore adopt $w{=}0.5$ as the default composition operating point in the main paper, where retention is near-additive without the severe overshoot observed at full strength.

\end{document}